\documentclass{article}
\usepackage{spconf,amsmath,graphicx}
\usepackage{url}
\usepackage{spconf,amsmath,graphicx}
\usepackage{amsfonts}
\usepackage{hyperref}
\usepackage{xcolor}
\usepackage{booktabs}
\hypersetup{
    colorlinks=false,
    linkcolor=gray,
    filecolor=magenta,      
    urlcolor=gray,
    }

\newcommand{\expval}[2]{{\mathbb E}_{#1}\left[ #2 \right]}
\def\D{{\mathcal{D}}}
\def\G{{\mathcal{G}}}
\title{SpeechPainter: Text-conditioned Speech Inpainting}
\name{Zal\'an Borsos$^*$\thanks{$^*$Correspondence to \texttt{zborsos@google.com}}, Matt Sharifi, Marco Tagliasacchi}
\address{Google Research}

\begin{document}
%
\maketitle
\begin{abstract}

We propose SpeechPainter, a model for filling in gaps of up to one second in speech samples by leveraging an auxiliary textual input. We demonstrate that the model performs  speech inpainting with the appropriate content, while maintaining speaker identity, prosody and recording environment conditions, and generalizing to unseen speakers. Our approach significantly  outperforms  baselines  constructed  using  adaptive TTS, as judged by human raters in side-by-side preference and MOS tests.

\end{abstract}
\begin{keywords}speech inpainting, multimodality, adversarial learning\end{keywords}

\section{Introduction}
\label{sec:intro}

\looseness -1 Applications involving recognition, enhancement, and recovery of signals in one modality can often benefit from input in other modalities. In the context of speech processing, there have been several efforts to separate, enhance or recover target speech sources based on visual and other sensory modalities. On the other hand, recovering speech based on text has received little attention, despite its wide range of use-cases. 

\looseness -1 In this work, we investigate text-conditioned speech inpainting, named by analogy with image inpainting, where a region of unknown pixels is inferred from the surrounding observed pixels~\cite{hays2007scene}. Our task is to infer a segment of missing speech up to one second long, given two inputs: i) the surrounding observed speech in an utterance, and ii) a transcript of the entire utterance, including both the missing and observed speech. In order to sound natural, the infilled audio should preserve speaker identity, prosody, recording environment conditions (reverb, background noise, bandpass filtering and other artifacts due to the recording microphone) and its speech content should correspond to the missing part according to the transcript. The approach should generalize to unseen speakers and content without any additional context provided besides the original short speech sample with the gap and the transcript. 

 Text-conditioned speech inpainting can be useful in several use-cases. For example, it could serve as a tool for users with speech impairments or language learners for fixing mispronounced words in recorded audio. In a similar manner, the model could be used for correcting grammar and word choice mistakes, as well as for restoring parts of speech affected by packet losses or excessive background noise.

\looseness -1 The problem of unconditional speech inpainting has been addressed in~\cite{6020748, bahat2015self}, with the main focus on packet loss concealment. These methods are only practical for short gap sizes (less than 250 ms). Relying on another modality for conditioning enables the inpainting of longer gaps. When conditioning on text, the speech inpainting approach of~\cite{7760374} can operate  with gaps up to 750 ms. However, the method requires a significant amount of parallel training data for each target speaker, which is impractical to collect in many settings. 

Recent advances in text-to-speech (TTS) systems~\cite{pmlr-v80-skerry-ryan18a,hsu2018hierarchical, pmlr-v139-min21b} demonstrate the capability of TTS to synthesize speech while transferring  speaker identity, prosody and some recording environment conditions from a short target speech sample. In fact, such approaches could already fulfill our requirements, provided that an additional module is designed for detecting the part of the synthesized speech corresponding to the gap, and ``stitching'' it back into the gap's place. We evaluate such a baseline in Section~\ref{sec:experiments}. Also relevant, but not directly related to our work, are methods that enhance speech based on modalities other than text: using facial crops of the target speaker, denoising~\cite{DBLP:journals/tetci/HouWLTCW18}, inpainting speech~\cite{av-speech-inpainting}  or separating a target speaker from a mixture~\cite{looking2listen, Wang2019}, and enhancing speech based on accelerometer data~\cite{seanet}, just to name a few.

\looseness -1 In this paper, we propose \emph{SpeechPainter}, a model capable of handling our challenging setup of text-conditioned speech inpainting. We demonstrate that the model can synthesize the missing speech content based on an \emph{unaligned} transcript, while maintaining speaker identity, prosody and recording environment conditions. We base our model on the Perceiver IO~\cite{jaegle2021perceiver}, a scalable architecture for multi-modal data that  demonstrated excellent performance and versatility over a range of tasks and input / output modalities. Another essential component for generating artifact-free speech is adversarial training with feature matching losses~\cite{pmlr-v48-larsen16}, commonly used in high-fidelity speech synthesis~\cite{ NEURIPS2019_6804c9bc, NEURIPS2020_c5d73680}. Our approach significantly outperforms baselines constructed using adaptive TTS, as measured by MOS of human raters. 
 We encourage the reader to listen to the samples produced by our model for unseen speakers and content.\footnote{\href{https://google-research.github.io/seanet/speechpainter/examples/}{https://google-research.github.io/seanet/speechpainter/examples/}}

\begin{figure*}[t]
\centering
\includegraphics[width=\textwidth]{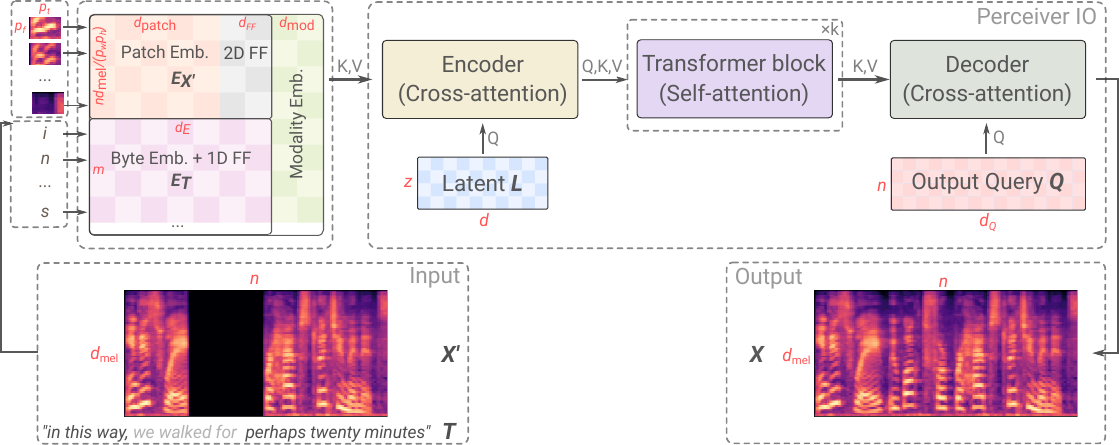}
\caption{\looseness -1  The architecture of SpeechPainter. The mel spectrogram of the audio input is split into patches, the text is split into characters. The concatenated embeddings are processed by Perceiver IO~\cite{jaegle2021perceiver}, which outputs inpainted mel spectrograms.\label{fig:speechpainter} } 
\end{figure*}

\section{Method}
\looseness -1 Before presenting our model, we first describe the requirements that motivate our design choices. First, the model should operate in the learning setup described in Section~\ref{sec:intro}, by filling in a gap of at most one second in a speech segment given the corresponding transcript. Second, in favor of simplicity and minimal effort for training data preparation, we assume that the transcript is not aligned with the corresponding speech sample. We allow for longer transcripts that have a subsequence covering the content of the speech sample with the gap. In this setup, the model should learn to discover the part of the transcript corresponding to the gap and synthesize its content accordingly. Training data preparation for this setup is trivial: take any speech dataset containing speech samples and transcripts (e.g., LibriTTS~\cite{libritts}, VCTK~\cite{yamagishi2019vctk}), randomly crop the speech samples to a few seconds, randomly create the gap and retain the whole original transcript. Third, for the sake of generality, we further require that the runtime and memory complexity of the model should scale well in terms of the length of the speech sample and transcripts, thus allowing the model to operate on longer sequences.

\subsection{Model} \label{subsec:model}
\looseness -1 We base our model architecture on Perceiver IO~\cite{jaegle2021perceiver}, which is capable of fulfilling the aforementioned requirements. In particular, Perceiver IO scales linearly in runtime and memory complexity with respect to the input and output sizes. This is achieved by processing the input using a cross-attention module with a learned latent query $L$ to a fixed-size latent space, which is then processed by a Transformer \cite{NIPS2017_3f5ee243}. Analogously, the output is constructed by cross-attending to the latent space using a learned output query $Q$. The choice of Perceiver IO as our base model, in contrast to autoregressive approaches, also allows us to use adversarial training (Section~\ref{sec:two-phase-training}), which we found to be crucial for removing robotic artifacts from the inpainted sequence.

\looseness -1 Instead of working with audio in the time domain, we choose the more compact log-scale mel spectrogram representation. This representation disregards phase information and emphasizes lower frequency details important for understanding speech, but also admits inversion algorithms capable of reproducing the audio in the time domain with high fidelity. In our experiments, we use a neural vocoder to perform the audio synthesis from log-mel spectrograms. The architecture of the vocoder is identical to MelGAN~\cite{NEURIPS2019_6804c9bc}, but it is trained with a multi-scale reconstruction loss as well as adversarial losses from both wave- and STFT-based discriminators, following~\cite{zeghidour2021soundstream}.

\looseness -1 The architecture of SpeechPainter is shown in Figure~\ref{fig:speechpainter}. The input and output of the model are $(X', T)$ and $X$, respectively, where $X \in \mathbb{R}^{n \times d_{\textup{mel}}}$ is the $n$-frame, $d_{\textup{mel}}$-band log-mel spectrogram of the target sample (without gap), $T$ is the transcript, $X':=\textup{mask}(X)$ and $\textup{mask}(\cdot)$ is an augmentation that masks consecutive frames of the spectrogram, creating a gap of random length covering at most one second, starting at a random frame. This is achieved by setting the corresponding subsequence of the target signal in the time domain to zero.

\looseness -1 The input keys / values to Perceiver IO's encoder (the first cross-attention) are constructed as follows. The log-mel spectrogram $X'$ is split into non-overlapping patches of size $p_t \times p_f$, and the resulting patches are embedded to $d_{\textup{patch}}$-dimensional representations through a learned linear layer after flattening. Then, the sequence of patches is concatenated with 2D Fourier positional encodings (FF, Fourier features)~\cite{NIPS2017_3f5ee243, pmlr-v139-jaegle21a} of dimensions $d_{FF}$, resulting in the spectrogram embedding $E_{X'} \in \mathbb{R}^{(n\cdot d_{\textup{mel}} / (p_t \cdot p_f) ) \times d_E}$, where $d_E =d_{\textup{patch}}+ d_{FF}$. For embedding the transcript,  $T$ is first padded with whitespaces to length $m$ (the maximum number of characters accepted by the model) in order to support batching, and, following~\cite{jaegle2021perceiver}, it is UTF-8 byte-embedded with embedding dimension of $d_E$ and is summed with Fourier positional encodings to obtain the text embedding $E_T \in \mathbb{R}^{m \times d_E}$. Finally, the learnable modality embeddings $e_{X}, e_T \in \mathbb{R}^{d_{\textup{mod}}}$ are appended to $E_{X'}$ and $E_T$, respectively, the concatenation of which is the key / value to the encoder's cross-attention.

\looseness -1 The learned latent $L \in \mathbb{R}^{z \times d}$ serves as the query to the encoder, the output of which is attended to by $k$ Transformer-style self-attention blocks. The task of the decoder is to produce the target log-mel spectrograms. To achieve this, the decoder's cross-attention receives as query a learned output query $Q \in \mathbb{R}^{n \times d_Q}$ and as key / value the output of the final self-attention layer. Finally, the output of the decoder is projected along the last dimension to match the dimension of the target spectrogram.

\subsection{Two-phase Training} \label{sec:two-phase-training}

We train the model in two consecutive phases in order to achieve both high signal reconstruction fidelity and high perceptual quality. In the first phase, we train the model with the $L_1$ reconstruction loss on log-mel spectrograms,
\begin{equation*}
    \mathcal{L}_\G^{\textup{rec}} = \frac{1}{nd_{\textup{mel}}} \expval{(X, T)}{\left\lVert X - \G(X', T) \right\rVert_1},
\end{equation*}
where $\G$ denotes the model. With this choice of loss function, the model learns to correctly identify and inpaint the content of the gap based on the transcript, maintaining speaker identity, prosody and recording environment conditions. However, the results contain robotic artifacts (see ablation in the accompanying samples). 

\begin{figure}[t]
\centering
\includegraphics[width=\columnwidth]{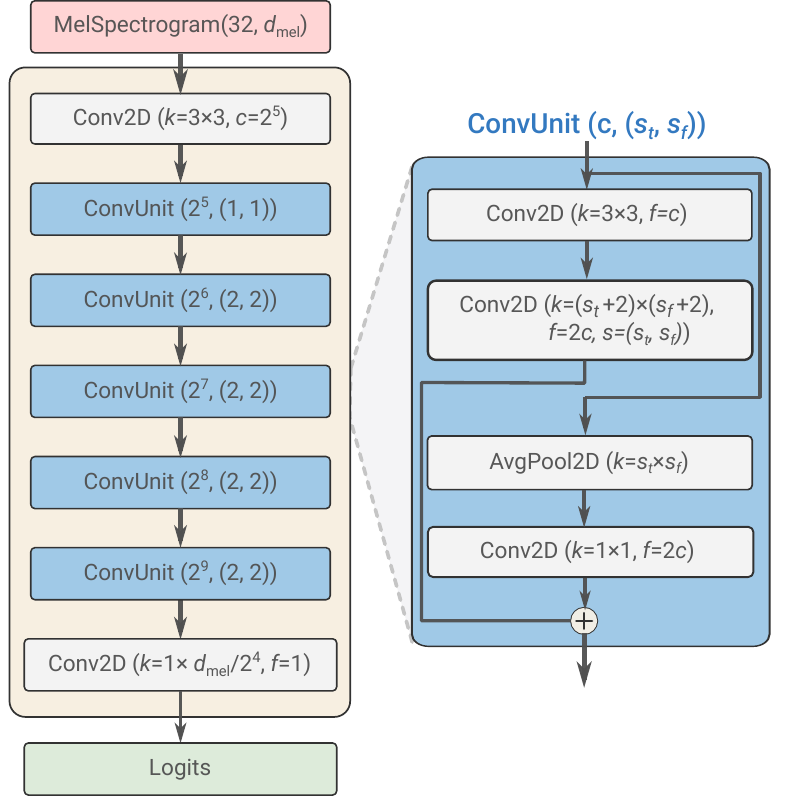}
\caption{The architecture of the discriminator operating on log-mel spectogram chunks. The convolution parameters \emph{k, f, s} denote the kernel size, number of filters and strides, respectively. The first two convolutions in each  block are preceded by ELU activation, and all convolution layers except the last are using layer normalization~\cite{ba2016layer}.} \vspace{-4mm}
\label{fig:mel-discriminator}
\end{figure}

\looseness -1 Therefore, we introduce a second training phase, where we perform adversarial training in order to achieve higher perceptual quality. We instantiate a discriminator $\D$ with the objective to distinguish between synthesized and real mel spectrogram chunks of 32 frames. At the same time, we optimize $\G$ to produce outputs that are indistinguishable from the ground truth by matching the feature representations in all layers of $\D$~\cite{pmlr-v48-larsen16, NEURIPS2019_6804c9bc}. $\D$ is a convolutional network closely resembling the single-scale STFT discriminator of~\cite{zeghidour2021soundstream}, shown in Figure~\ref{fig:mel-discriminator} --- for further details about the architecture, we refer the reader to~\cite{zeghidour2021soundstream}. The discriminator is trained with the hinge loss,
\begin{equation*}
\begin{split}
    \mathcal{L}_\D = \, &  \expval{(X, T)}{\sum_t \max \left(0, 1-\D_t(X)\right)}  + \\ 
    & \expval{(X, T)}{\sum_t \max \left(0, 1+\D_t(\G(X', T))\right)},
\end{split}
\end{equation*}
where $\D_t$ is the $t$-th logit of the discriminator along the time axis. The feature matching loss for $\G$ is defined as
\begin{equation*}
    \mathcal{L}_\G^{\textup{feat}} \!=\!  \expval{(X, T)}{\frac{1}{\ell}\sum_{i=1}^\ell\!\frac{1}{d_i}\left\lVert \D^{i}\!(X) \!-\!  \D^{i}(\G(X', T))\right\rVert_1}
\end{equation*}
where $\ell$ is the number of layers of $\D$,  $\D^i$ denotes the discriminator features at layer $i$ and $d_i$ is the dimension of $\D^{i}$. In the second phase of the training, $\G$ is optimized using the loss $\mathcal{L}_\G^{\textup{rec}} + \lambda_\textup{feat} \mathcal{L}_\G^{\textup{feat}}$ with $\lambda_\textup{feat}=10$.

\section{Experiments} \label{sec:experiments}
 \looseness -1 We train our model from scratch on LibriTTS at 24kHz, without relying on transfer learning or reusing pre-trained embeddings. We choose the target audio length equal to three seconds, randomly cropped from the target utterance, which, in the case of LibriTTS, corresponds to an entire sentence. We feed the transcript of the whole target utterance to the model, and require the model to learn to determine the content of the gap based on the input audio containing the gap, thus minimizing the data preparation effort. Random amplitude scaling is used as the sole data augmentation.
 
\begin{table*}[t]
\centering 
\caption{Side-by-side preference and MOS tests (mean and standard error of the scores). In both tests, SpeechPainter outperforms adaptive TTS. According to aggregated MOS, human raters can detect the artifacts of inpainting with SpeechPainter. \label{table:res}}

\begin{tabular}{cc|ccc}
\toprule
                     & \textbf{Preference}   & \multicolumn{3}{c}{\textbf{MOS}} \\
\multicolumn{1}{l}{} & Speechpainter vs. TTS & TTS   & SpeechPainter  & Original (Target)  \\ \midrule
LibriTTS clean       & 0.72 $\pm$ 0.10       &  3.06 $\pm$ 0.10   & 3.60 $\pm$ 0.09  & 3.68 $\pm$ 0.09 \\
LibriTTS other       & 0.55 $\pm$ 0.12       & 2.77 $\pm$ 0.12 & 3.12 $\pm$ 0.10 &   3.30 $\pm$ 0.12 \\
VCTK                 & 0.80 $\pm$ 0.10       & 3.38 $\pm$ 0.10 &  3.70 $\pm$ 0.09  & 3.98 $\pm$ 0.08  \\ \midrule
Aggregated           & 0.69 $\pm$ 0.06       & 3.07 $\pm$ 0.06  &  3.48 $\pm$ 0.06 & 3.66 $\pm$ 0.06 \\ \bottomrule
\end{tabular}
\end{table*}

\looseness -1 We operate on log-scale mel spectrograms with hop length of 12.5 ms, frame length of 40 ms and $d_\textup{mel}=64$ channels, resulting in $n=240$ dimensional log-mel spectrogram $X$ along the time axis. We divide the spectrogram into non-overlapping patches of $p_t=8$ by $p_f=4$. The spectrogram patches are mapped to $d_\textup{patch}=768$ dimensions after flattening, and are concatenated with 2D Fourier positional encodings with $d_{FF}=256$ to obtain $d_E = 1024$-dimensional embeddings for each patch. We pad the text to $m=500$ characters and use modality embeddings of $d_\textup{mod}=16$ dimensions. The learned latent dimensions are determined by $z=512$ and $d=1024$, while for the learned decoder query we have $n=240$ and $d_Q=512$. We use $k=16$ self-attention layers, while both the cross- and self-attention layers have $h=16$ heads. The query, key and value dimensions for all attention layers are set to $d/h = 64$. In the attention blocks, we use two-layer feed forward networks with hidden layer dimension of $d=1024$. These parameter choices lead to a model with 112M parameters.
In the first training phase, we train the model with Adam~\cite{DBLP:journals/corr/KingmaB14} with batch size of 256 for 3M steps, using learning rate of $10^{-4}$ decayed by a factor of 5 after every 800K steps. The second phase consists of 100K steps, where both the model and the discriminator are trained with Adam using a learning rate of  $5 \cdot 10^{-5}$.  

We evaluate our method for unseen speakers and content using samples from LibriTTS test-clean and test-other splits, as well samples from VCTK. We first trim the leading and trailing silence, and retain samples that are at least three seconds long after trimming. From these, we randomly choose 20 samples from each split (60 samples in total) from different speakers and randomly crop them to three seconds. We mask out a sequence of length between 750 ms and 1000 ms starting at a random location, while ensuring that the masked-out part has at least 300 ms left and right context. 

As we are unaware of existing methods for text-conditioned speech inpainting suitable for our learning setup described in Section \ref{sec:intro}, we design one based on adaptive TTS. For this, we synthesize the target transcript using a TTS system capable of transferring the speech style from the short target sample (three seconds, without masking), identify the part corresponding to the masked-out sequence, and perform the inpainting by inserting the identified part using a fade-in / fade-out technique. We rely on Meta-StyleSpeech~\cite{pmlr-v139-min21b} for adaptive TTS, using the pretrained model provided by the authors\footnote{\href{https://github.com/KevinMIN95/StyleSpeech}{https://github.com/KevinMIN95/StyleSpeech, accessed on 21.11.2021}} and a vocoder matching Meta-StyleSpeech's frontend trained in the same manner as SpeechPainter's vocoder (Section \ref{subsec:model}).

\looseness -1 We conduct side-by-side preference tests and Mean Opinion Score (MOS) tests with 6 human raters for assessing the naturalness of the inpainted samples. The raters were provided with the definition of a natural speech sample as a sample containing intelligible speech with no breaks in speaker identity, prosody and recording conditions. In the preference test, the raters are provided with the 60 pairs of samples corresponding to inpainting with TTS and SpeechPainter. The raters are asked to provide an score in $\{-2, -1, 0, 1, 2\}$, where a positive score indicates that the first sample is more natural than the second one. For MOS, we ask the raters for the naturalness on a five-point Likert scale (1-Bad, 2-Poor, 3-Fair, 4-Good, 5-Excellent) of the $60\cdot3$ samples corresponding to the target samples, inpainting via TTS and SpeechPainter, in a random order. The results in Table \ref{table:res} show the scores over splits, as well as the aggregated scores. Although not designed for our use-case, the approach based on adaptive TTS represents a strong baseline according to the MOS tests. The superiority of SpeechPainter over adaptive TTS is expressed both in side-by-side preference test and in MOS scores on individual splits, as well as in the aggregated scores. According to aggregated MOS, human raters can also detect the artifacts of inpainting, and express a statistically significant preference of real samples over the ones inpainted with SpeechPainter. 

 \looseness -1  \textbf{Limitations.} We experimented with real recordings in order to identify the limitations of the current model. We found that the model struggles in generalizing to recording environment conditions that significantly deviate from the ones encountered during training. The model fails to produce intelligible speech for highly reverberant, noisy samples, or when the recording microphone introduces noticeable distortions. Similarly, the model fails to generalize to unseen accents and speech tempos. There is also a degradation in speaker identity preservation for longer gap sizes, as well as a tendency towards producing a monotone prosody. We demonstrate the failure cases in our accompanying samples.

\section{Conclusion}

We formulated speech inpainting in the setup where only a short speech sample containing a gap and its corresponding transcript are available. We presented SpeechPainter, a model capable of synthesizing the missing speech content based on its unaligned transcript. We demonstrated that the model is capable of generalizing to unseen speakers and content, while maintaining speaker identity, prosody and recording environment conditions.

\section{Broader Impact}

In Section~\ref{sec:intro}, we discussed several socially beneficial use-cases of text-conditioned speech inpainting. The main assumption was that the target speaker corresponds to the user of the model and the speech inpainting process happens with the consent of the speaker. However, since the model generalizes to unseen speakers and content, it raises the potential of third-party abuse of altering content and possibly the message of the speech of an unknown speaker without their  consent; this could be used to spread misinformation or for phishing attacks. Therefore, crucial components that should accompany such models are protocols for ensuring the speaker's consent and the disclosure of the performed modifications, as well as systems for detecting inpainted speech sequences.

\looseness -1 From the perspective of fairness, the model reflects the biases present in its training data. The most obvious manifestation of such a bias when training on LibriTTS (where native EN-US speakers outnumber other groups) is that accents might not be preserved: for example, the accent in the inpainted sequence for a non-native speaker is likely to be altered to an EN-US accent. A possible mitigation of this issue is better representation of a wide range of accents in the training data.

\section{Acknowledgements}
We would like to thank John Hershey and Johnny Soraker for their insightful feedback on this work.

\bibliographystyle{abbrv}
\bibliography{refs}

\end{document}